\documentclass[fleqn,10pt]{wlscirep}
\usepackage[utf8]{inputenc}
\usepackage[T1]{fontenc}
\usepackage{caption}
\usepackage{subcaption}
\title{An On-Chip Continuous Wave Terahertz Spectrometer}

\author[1,*]{James Seddon}
\author[1+]{Chris Graham}
\author[1+]{Marie Georgiades}
\author[1+]{Cyril Renaud}
\author[1+]{Alywn Seeds}
\affil[1]{Department of Electronic and Electrical Engineering, University College London, London, WC1E 7JE, U.K}

\affil[*]{james.seddon@ucl.ac.uk}

\affil[+]{these authors contributed equally to this work}

\begin{abstract}
Continuous Wave (CW) Terahertz spectroscopy enabled by photomixing is a promising high precision spectroscopic tool for the examination of a wide variety of samples including biological, chemical, and solid state. However, often it would be of interest to examine isolated samples free from bulk effects that broaden spectral features. In this form samples with low concentrations of absorbers have a reduced cross section making coupling to an external driving field a challenge. By utilising THz metamaterials one can borrow the concepts of surface enhanced Raman spectroscopy to confine and concentrate the THz fields to increase sensitivity for samples with a small cross section. Our work combines high-speed photodiode technology and THz metamaterials to offer a solution to this challenge. We present here a proof of concept on-chip THz spectrometer integrated with a metamaterial waveguide.
\end{abstract}
\begin{document}

\flushbottom
\maketitle
% * <john.hammersley@gmail.com> 2015-02-09T12:07:31.197Z:
%
%  Click the title above to edit the author information and abstract
%
\thispagestyle{empty}

\section*{Introduction}

The Terahertz (THz) spectral region between 0.1 THz and 10 THz\cite{Dhillon2017-jk} is of great interest due to the wide variety of spectral absorption signatures. These range from vibrational\cite{Globus2013-vw} and rotational dynamics\cite{Motiyenko2014-qw} of molecular systems, and molecular conformal responses caused by hydration\cite{Wirtz2018-an,Esser2018-jt} and bonding\cite{Swearer2018-bz}, to crystal field splitting in rare earth doped laser crystals\cite{Matmon2016-xb} and inter-sub-band transitions in semiconductor quantum wells\cite{Heyman1998-rf}. 
To date this region of the electromagnetic spectrum has been predominantly explored using THz time domain spectroscopy (TDS), with several commercially available solutions developed\cite{Stanze2011-ay}. 

The main drawbacks of THz TDS spectroscopy are its bulk and expense, due to its reliance on a femtosecond pulsed laser system, and the spectral resolution, limited by the source stability and length of the optical delay line. An alternative system for THz spectroscopy is frequency domain continuous wave (CW) spectroscopy. CW spectroscopy involves the tuning of a heterodyne beat note between two telecommunications (1550 nm) band lasers to produce tuneable THz radiation in high speed photomixers. One such photomixing source, the Uni-travelling carrier photodiode\cite{Ishibashi1997-gn} (UTC-PD) has shown promise in terms of output power levels across the THz region between 0.1 and 3 THz, reaching milliwatt levels of output power around 0.1 THz and microwatt levels around 1 THz. 

In addition to the performance as a THz source the UTC-PD has also been demonstrated as an optically pumped mixer\cite{Nagatsuma2010-ez,Rouvalis2011-fk,Rouvalis2012-zw} for coherent detection of THz waves. While the conversion losses of the optically pumped mixers are high compared to other optically pumped mixers such as the InGaAs photoconductor, it presents the opportunity for a THz source and detector to be monolithically integrated on the same wafer to create a compact THz spectrometer.  Conventional THz spectrometer systems typically rely on antennas either in the form of planar substrate integrated antennas or coupled from packaged rectangular waveguide modules to radiate THz signals into a beam path formed from off-axis parabolic mirrors\cite{Stanze2011-ay,TOPTICA_Photonics2021-ep}. Samples are then placed at points either in the collimated beam path or at a focal point of the mirror. 

For samples that have a small absorption cross section or have high attenuation coefficients, such as molecular samples in an aqueous environment, detection of spectral signatures can be challenging due to the strong attenuation of THz radiation by water. In the case of aqueous solutions, the interaction volume between the THz signal and sample can be greatly reduced to mitigate the effects of absorption in water, however the reduced interaction volume can reduce the magnitude of the power absorbed as the measurable absorption is inversely proportional to sample thickness\cite{Naftaly2013-is}, requiring systems with very high dynamic range (DR) and signal to noise ratio (SNR) to resolve the weaker spectral signatures. A solution to this is to confine and concentrate the THz radiation around the small volume samples.

This has been demonstrated for THz TDS spectrometers using both metasurfaces and the evanescent fields around on-chip planar waveguides\cite{Byrne2008-zo}. 
By using planar waveguides, the source and detector elements can be integrated together with the waveguide to create a compact on-chip spectrometer platform. Furthermore, individual resonator elements can be evanescently coupled to the waveguide to provide greater field enhancement and localisation. The highly confined fields of the resonators can then be used to probe nonlinear physical phenomena such as the strong coupling regime and electronically induced transparency.

\begin{figure}[ht]
\centering
\includegraphics[width=\linewidth]{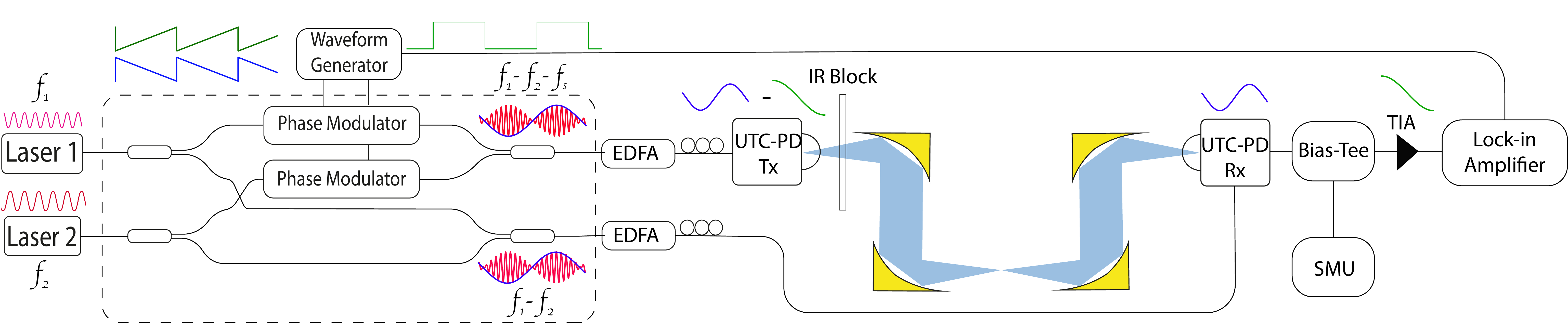}
\caption{Schematic drawing of the free space UTC-PD spectrometer used to evaluate the optimum optoelectronic mixing parameters and dynamic range of the UTC-PD as a receiver element (Rx). The spectrometer relies on two free running laser sources which are used to generate a heterodyne beat note in the transmitter (Tx) UTC-PD. A differential sawtooth wave is used to drive two phase modulators to apply an up shift and a down shift to laser 1 and 2 respectively. This frequency offset between the Tx and Rx allows coherent detection of the incoming THz waves to recover amplitude and phase information of samples placed in the THz beam path. A transimpedance amplifier with 104 A/W gain is placed after the Rx UTC-PD and a lock-in amplifier is used to record the amplitude and phase of the down converted THz signals with a 100 ms time constant. }
\label{fig:FS_Schematic}
\end{figure}

Initial demonstrations of THz on-chip systems consisted of microstrip transmission lines deposited on amorphous silicon photoconductive sources\cite{Auston1980-ao} and co-planar waveguide (CPW) transmission lines on silicon on sapphire substrates\cite{Grischkowsky2000-fc}. Microstrip waveguides integrated with low temperature Gallium Arsenide (LT-GaAs) have also been demonstrated as on chip TDS systems using flip chip bonding techniques\cite{Byrne2008-zo} and operation at cryogenic temperatures\cite{Wood2006-at}. The on-chip spectrometers were used to measure both the spectrum of polycrystalline lactose\cite{Byrne2008-zo} for measurement of permittivity and loss tangent\cite{Park2020-nd} and integrated with a microfluidic chip to determine the Debye relaxation parameters of liquid samples\cite{Swithenbank2017-dd}. CPWs have also been integrated with an LT-GaAs system\cite{Wood2013-ip} albeit with reduced bandwidth due to losses from leakage of the propagating mode into the substrate. Single conductor Goubau lines\cite{Goubau1951-nc} have been demonstrated as on-chip THz waveguides integrated with both LT-GaAs systems\cite{Dazhang2009-lm} and UTC-PDs\cite{Akalin2019-oz}. Goubau lines also have issues with losses due to substrate modes and issues with dispersion in TDS systems. Substrate losses can be mitigated by thinning the dielectric substrate or by optimising the input transition to the Goubau line\cite{Russell2013-xu}.

An alternative single conductor waveguide is the spoof plasmon polariton (SPP) waveguide which is formed from a periodic array of grooves in a metallic surface, initially proposed by Goubau\cite{Goubau1950-zq} as a textured surface to a metal wire for guiding of surface waves. Barlow and Cullen extended this concept to planar metallic surfaces\cite{Barlow1953-xx}. This was further developed by Pendry et al\cite{Pendry2004-zx} to describe anomalous transmission through subwavelength metallic hole arrays. A planar form of the SPP waveguide was demonstrated\cite{Shen2013-is} by reducing the thickness of the grooved surface to nearly zero enabling deposition of planar SPP compatible with existing semiconductor device fabrication techniques. 

The SPP waveguides can be engineered to exist for microwave, sub-millimetre and THz frequencies\cite{Williams2008-pr} based on engineering of the groove geometry, which allows control over the dispersion relation of the waveguide, this can help to mitigate radiative losses into substrate modes. One of the primary limitations of the SPP waveguide is the increase in Ohmic losses close to the spoof plasma frequency where the propagating wave velocity is slowed and tightly confined to the textured surface. This is predominantly an issue in the THz spectral range where metal losses are larger than in the microwave spectral range. 

Despite the increased losses the tightly confined fields convey an additional benefit of increased sensitivity to changes in the local dielectric permittivity which enables applications such as biomolecule sensing and spectroscopy in small volume droplets. Waveguide integrated bio-sensing has been previously demonstrated\cite{Nagel2002-tf} where a microstrip filter was used for detection of proteins and DNA. 
The principle of operation of these biosensors is to detect a frequency shift in the stop band frequency of the filter. Many of these devices are demonstrated using free space THz radiation or by coupling to on-chip waveguides via RF probes or antennas. Moving to an on-chip modality will reduce the footprint of such devices, however there have been few examples of sensing waveguides integrated with active elements such as sources or detector. UTC-PDs integrated with distributed feedback lasers, modulators and SOAs have previously been demonstrated\cite{Lamponi2013-sc} along with on chip integration of the self-heterodyne spectrometer\cite{Lo2017-wb} on an Indium Phosphide (InP) platform and on Silicon\cite{Nishi2014-un}. These studies demonstrate the potential for monolithic integration of all elements of the spectrometer system, which would enable development of a handheld spectroscopy and sensing platform.

In this work we demonstrate a proof of concept on chip spectrometer based on monolithically integrated UTC-PD emitters and receivers and a metamaterial waveguide. The on-chip spectrometer is the first monolithic CW THz spectrometer to demonstrate > 40 dB of dynamic range and an on-chip bandwidth of > 300 GHz. 

\section*{Results}

\subsection*{UTC-PDs as emitters and receivers}
Previous studies have examined the UTC-PD as an optically pumped THz receiver. Receiver characteristics were evaluated in a square law operation and a homodyne detection regime\cite{Nagatsuma2010-ez}. The optimum mixing point of operation was observed at moderate photocurrents of 4 mA and bias voltages of -1 V with the highest sensitivity observed for the homodyne detection scheme. Similar investigations into travelling wave UTC-PDs have also been carried out\cite{Rouvalis2011-fk}. The optimum mixing point was observed at higher photocurrents and bias voltages respectively. 
\begin{figure}[ht]
    \centering
    \includegraphics[width=0.6\linewidth]{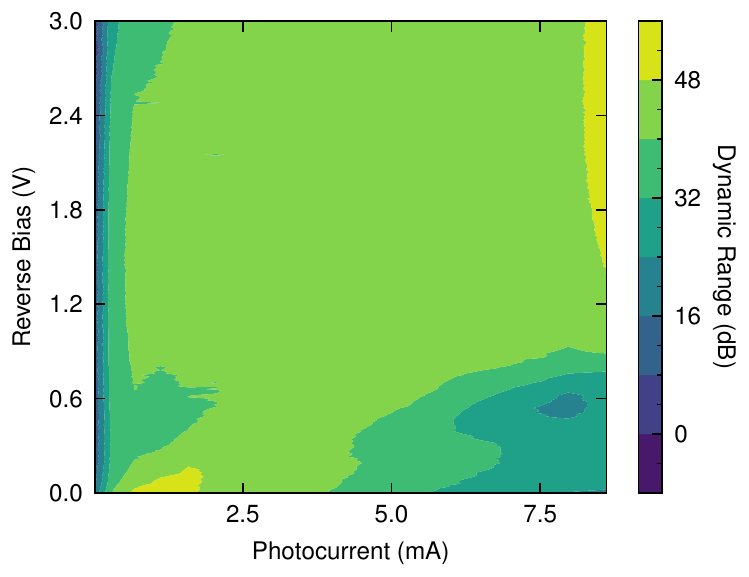}
    \caption{Colourmap of the bias voltage and Rx photocurrent which is proportional to optical drive power vs dynamic range at 100 GHz showing the optimum point of operation for the UTC-PD receiver at high bias and high optical power.}
    \label{fig:FS-DR-Colourmap}
\end{figure}
A self-heterodyne\cite{Hisatake2013-tp,Kim2013-by} spectrometer has been demonstrate to evaluate the performance of the UTC-PD as a mixer as part of a photodiode based CW THz spectrometer. Again, demonstrating the lowest conversion losses for heterodyne detection at a photocurrent of 4 mA and bias of -1 V. This is reported as the point at which the LO power of the optoelectronic mixer is expected to saturate with the predominant mixing mechanism being the result of a dynamic capacitance which results from charge storage in the photo-absorption layer\cite{Fushimi2004-ah}. The signal received at the receiver UTC-PD can be expressed by\cite{Hisatake2013-tp,Hisatake2014-np}:
\begin{equation}
    E_{RF}(t) \propto T(\omega_{RF})\cos(\omega_{RF}t+\phi(\omega_{RF})\varphi_{2n}(t)-\varphi_{1n}(t))
    \label{eqn:self-het-1}
\end{equation}
where $\omega_{RF}$ is the heterodyne beat frequency with the additional Serrodyne\cite{Johnson1988-oe,Poberezhskiy2005-js} frequency shift $2\pi\left(f_1-f_2-f_s\right), T\left(\omega\right)$ and $ \phi\left(\omega\right)$ are the amplitude and phase of incoming terahertz wave carrying the spectroscopic information.
The phase noise of the two free running lasers are given by $\varphi_{1n}\left(t\right) and \varphi_{2n}\left(t\right)$ respectively\cite{Hisatake2014-np}. The local oscillator (LO) signal driving the receiver UTC-PD is given by:
\begin{equation}
    E_{LO}(t)\ \propto\ \cos{\left(\omega_{LO}t+\omega_{LO}\tau+\ \varphi_{2n}(t+\tau)-\varphi_{1n}(t+\tau)\right)}
    \label{eqn:Self-het-2}
\end{equation}
where $\omega_{LO}$ is the un-shifted beat note between laser 1 and laser 2 $2\pi\left(f_f-f_2\right)$ and $\tau$ is the variable path delay between the LO and RF paths. The frequency translation is applied to the transmitter path to reduce the effect of parasitic amplitude modulation from the phase modulator\cite{Kim2013-by}. By moving this to the source side, the magnitude of the transmitted parasitic beat note is reduced. The down converted IF signal is given by:

\begin{equation}
    i_{IF}\propto T\left(\omega_{RF}\right)\cos{\left(\omega_st+\phi\left(\omega_{RF}\right)-\omega_{LO}\tau+\Delta\varphi_{2n}\left(t\right)-\Delta\varphi_{1n}\left(t\right)\right)}
    \label{eqn:self-het-3}
\end{equation}

where $\omega_s=2\pi f_s$ and $\Delta\varphi_{1n}\left(t\right)=\varphi_{1n}\left(t\right)-\varphi_{1n}\left(t+\tau\right), \Delta\varphi_{2n}\left(t\right)=\varphi_{2n}\left(t\right)-\varphi_{2n}\left(t+\tau\right).$ Using dual phase lock-in detection the in phase $ and quadrature data $ is measured as:

\begin{equation}
    I\left(\omega_{RF}\right)=T\left(\omega_{RF}\right)\cos{\left(\phi\left(\omega_{RF}\right)+\varphi_n\right)}
    \label{eqn:self-het-4}
\end{equation}

\begin{equation}
    Q\left(\omega_{RF}\right)=T\left(\omega_{RF}\right)\sin{\left(\phi\left(\omega_{RF}\right)+\varphi_n\right)}
    \label{eqn:self-het-5}
\end{equation}

Where $\varphi_n$ is the time averaged excess phase noise arising from fluctuations in the optical path length of the fibre interferometer indicated within the dashed box in  Figure \ref{fig:FS_Schematic}. This was controlled by encasing the fibre interferometer inside an insulated box to minimise temperature fluctuations. 

\begin{figure}[ht]
    \centering
    \includegraphics[width=0.6\linewidth]{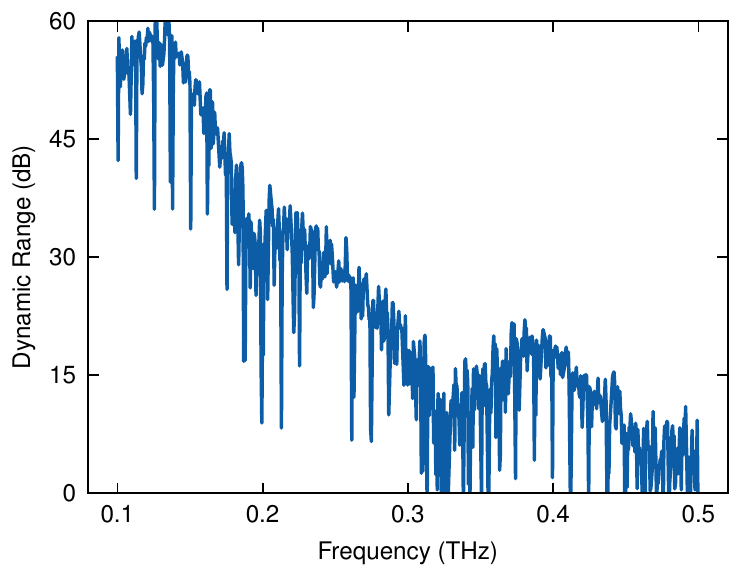}
    \caption{ Dynamic range spectrum for an antenna integrated UTC-PD spectrometer. The sharp regularly spaced dips in the spectrum are due to mode hopping of the tuneable laser as its wavelength is swept. The roll-off of dynamic range is due to the frequency response of the transmitter photodiode}
    \label{fig:FS-DR-Spectrum}
\end{figure}

Antenna integrated UTC-PD modules were used to evaluate the optimum point for UTC-PD receiver operation. A schematic diagram of the experimental arrangement is shown in Figure \ref{fig:FS_Schematic}. The bias voltage to the receiver UTC-PD was swept between 0 and -3V while recording the average magnitude of the lock-in amplifier voltage and DC photocurrent for increasing optical drive power. The same sweep was also performed with the THz beam blocked and the RMS noise measured to determine the dynamic range of the system\cite{Naftaly2009-ef}. Figure \ref{fig:FS-DR-Colourmap} shows the dynamic range increasing with reverse bias for the receiver for a range of increasing LO optical pump powers. The point at which the highest magnitude IF power is recorded is at the highest optical power and a bias of -3 V. There is a second peak at low optical LO drive powers at 0V bias also. 

Upon establishing the optimum mixing point the wavelength of laser 2 was tuned to generate a heterodyne frequency between 0.1-0.5 THz while the magnitude of the lock-in voltage was recorded. Figure \ref{fig:FS-DR-Spectrum} shows the dynamic range spectrum of the system. with an 80 kHz IF frequency. The peak dynamic range of 60 dB is observed around 150 GHz. This peak is consistent with previous measurement of the output power of the  transmitter UTC-PD\cite{Seddon2022-mz}. 

\subsection*{UTC-PD Impedance }
 To maximise power transfer in a transmission line to a load impedance, its impedance should be matched to the transmission line and source impedance. Typically, in most test equipment this impedance is 50 Ohms. For a UTC-PD it has been shown previously that the impedance is a complex frequency dependant value. An equivalent circuit model\cite{Natrella2016-eu} was developed to model the UTC-PD impedance for complex conjugate matching to an integrated antenna. For the lab on chip integration with metamaterial waveguides the source and load impedance of the UTC-PDs are connected via a waveguide with a defined characteristic impedance. To model accurately the power transfer between source and receiver UTC-PD it is necessary to develop a model for the impedance of both the source and receiver in their respective operating conditions.
 
 \begin{figure}[ht]
    \centering
    \includegraphics[width=0.6\linewidth]{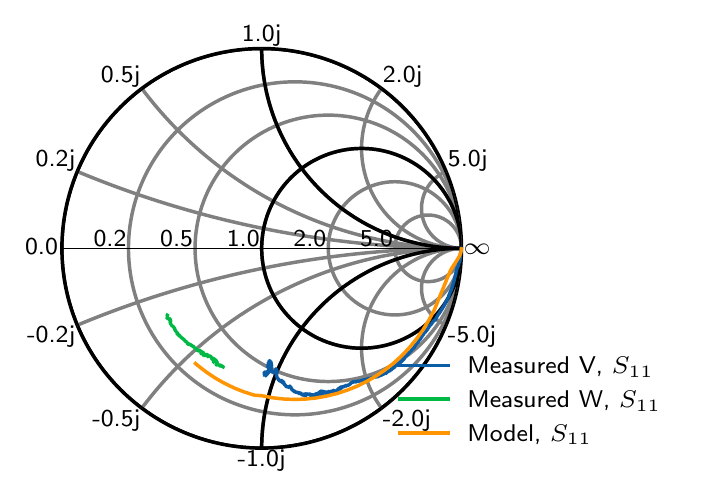}
    \caption{Measurement of the receiver photodiode impedance in two frequency bands 0.1-67 GHz (V) and 75-110 GHz (W) at the operating conditions determined in the free space spectrometer. A model impedance based on an equivalent circuit model shows good agreement with the measured impedance data. }
    \label{fig:UTC-PD-Smith}
\end{figure}
 
Based on the optimum mixing point determined in the previous section, the impedance of a co-planar waveguide integrated UTC-PD was measured from 0.1 to 67 GHz and in the W band (75-110 GHz) using a VNA with frequency extenders for the W band measurements.
The impedance measurement was carried out at -3V at 20 dBm optical input power to best represent the impedance at the optimum operating conditions of the receiver. Figure \ref{fig:UTC-PD-Smith} shows the impedance of the UTC-PD receiver with the impedance of an equivalent circuit model\cite{Natrella2016-eu}. There is good agreement between the model impedance and the measured impedance across both frequency bands. The UTC-PD circuit model parameters can be converted into effective material conductivity and permittivity for the associated resistance and capacitance of the RC circuits respectively. These materials are then used in CST Microwave studio to generate a full wave model of the device impedance which can then be used to design the metamaterial interconnecting waveguide.

\subsection*{Metamaterial waveguide design and integration}
\begin{figure}
     \centering
     \begin{subfigure}[b]{0.4\textwidth}
         \centering
         \includegraphics[width=\textwidth]{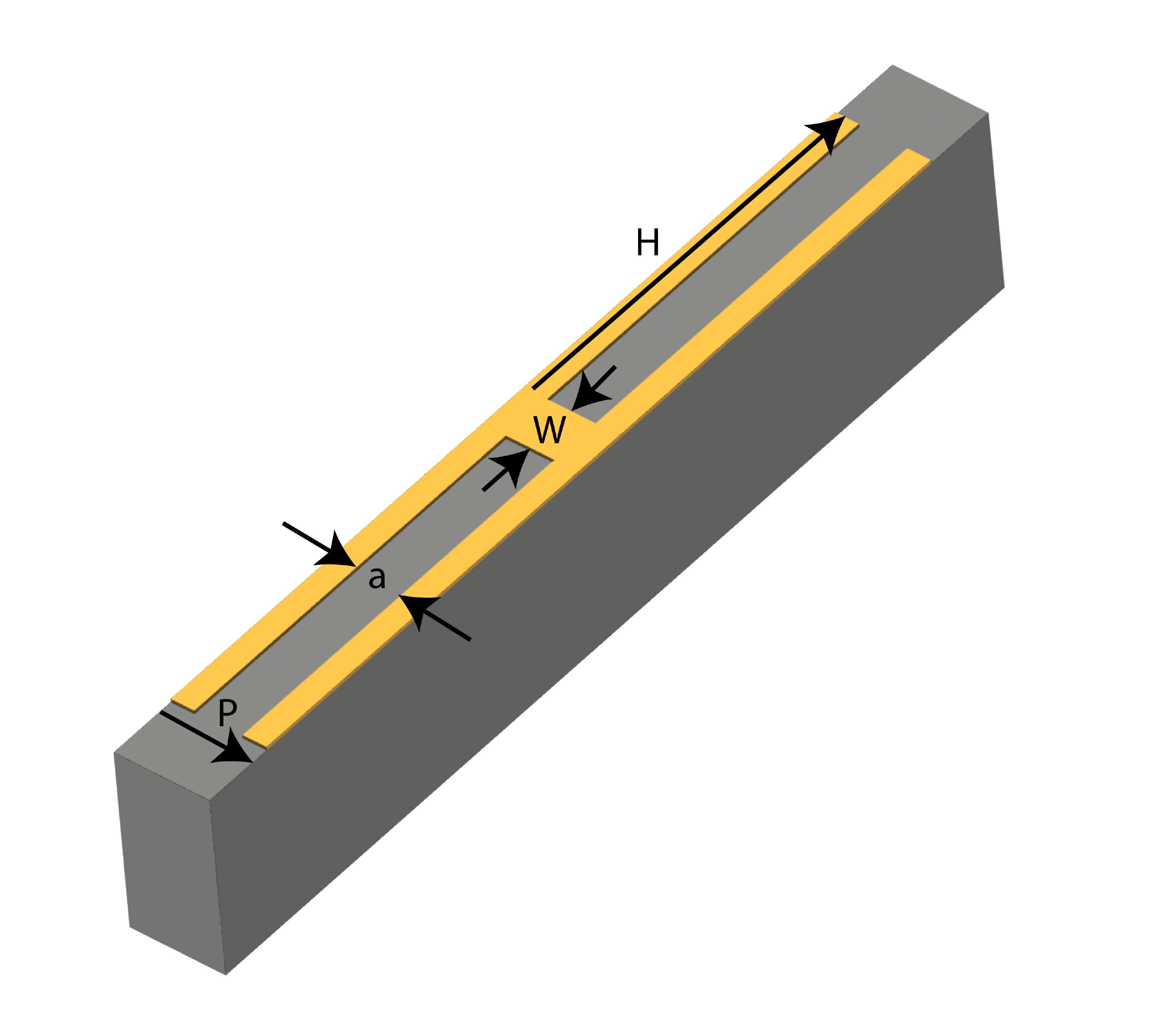}
         \caption{}
         \label{fig:SPP-Unit-Cell}
     \end{subfigure}
     \hfill
     \begin{subfigure}[b]{0.5\textwidth}
         \centering
         \includegraphics[width=\textwidth]{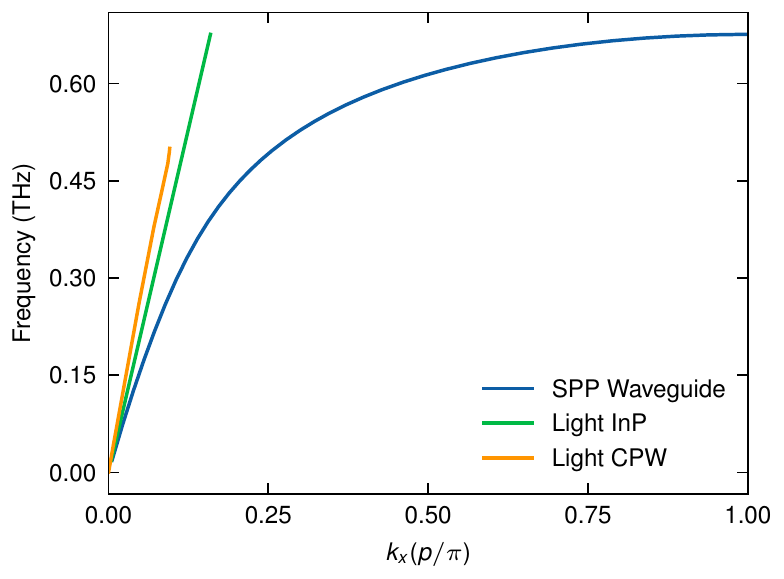}
         \caption{}
         \label{fig:SPP-Dispersion}
     \end{subfigure}
     \hfill
     \caption{\ref{fig:SPP-Unit-Cell} Schematic drawing of the SPP waveguide unit cell. The spoof plasma frequency is determined by the length of the grooves (H) the period (P) of the waveguide was set at 10 $\mu m$ with a duty cycle (a) of 0.5P. \ref{fig:SPP-Dispersion} The numerically simulated dispersion diagram for the unit cell in a is shown along side the light line in InP and a 50 $\Omega$ CPW waveguide. The dispersion curve of the CPW waveguide lies above the light line while the SPP waveguide lies below it and asymptotically tends towards the spoof plasma frequency }
\end{figure}

The design parameters for spoof plasmon polariton (SPP) waveguides are the period, groove height and duty cycle, along with the width of the centre conductor line detailed in Figure \ref{fig:SPP-Unit-Cell}. One of the key issues with planar waveguides on high permittivity materials such as InP and GaAs is radiative losses due to substrate modes\cite{Rutledge1983-du}. The dispersion diagram in Figure \ref{fig:SPP-Dispersion}. shows the numerically simulated dispersion of a SPP waveguide compared to a CPW waveguide on InP. 

The dispersion relation is shown for the quasi transverse electromagnetic CPW mode based on a 10 $\mu$m wide signal and 9 $\mu$m gap line at a 50 Ohm characteristic impedance. The phase velocity in the CPW line is greater than that of a wave travelling in the InP substrate, which leads to generation of a Cherenkov like shock wave cone which radiates the guided waves into the substrate\cite{Grischkowsky2000-fc,Rutledge1983-du,Cao2014-dh}. The SPP waveguide can be considered as a slow-wave type structure with a slower phase velocity than the light  line in InP $(2\pi/\lambda_{InP})$ which asymptotes to 0 as the spoof plasma frequency is approached. The expected cut-off frequency of the waveguide with a 60 $\mu m$ groove height can be seen from the dispersion diagram of Figure \ref{fig:SPP-Dispersion} given by the spoof plasma frequency $\omega_p=2\pi c/\left(2h\right)$ where h is the groove height\cite{Huidobro2018-kp} as 630 GHz. 

\begin{figure}
     \centering
     \begin{subfigure}[b]{0.45\textwidth}
         \centering
         \includegraphics[width=\textwidth]{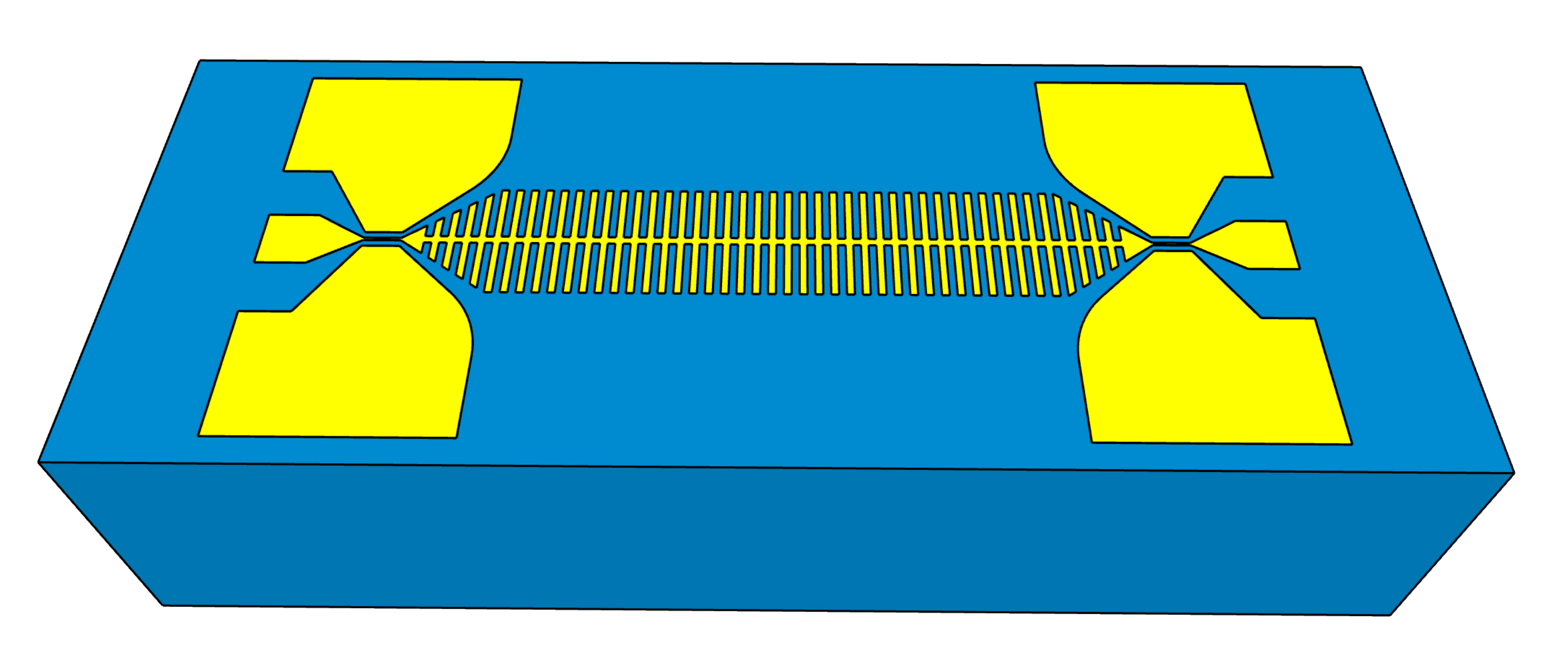}
         \caption{}
         \label{fig:SPP-Waveguide}
     \end{subfigure}
     \hfill
     \begin{subfigure}[b]{0.5\textwidth}
         \centering
         \includegraphics[width=\textwidth]{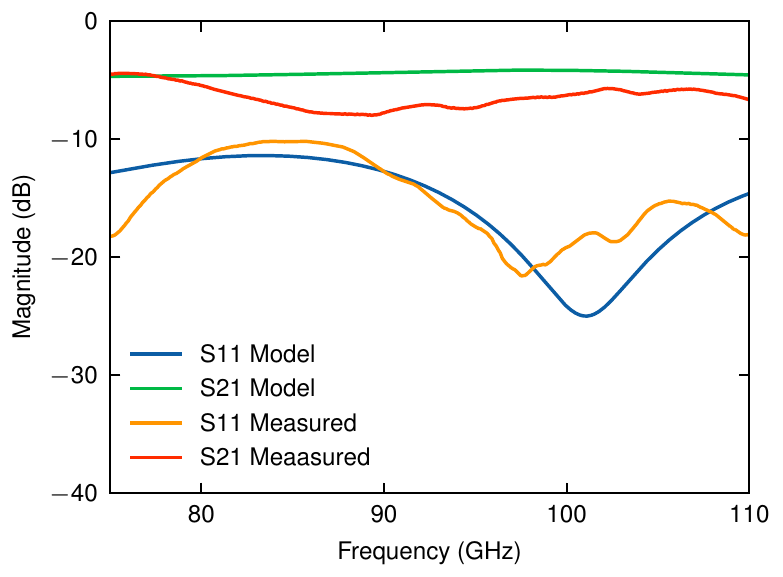}
         \caption{}
         \label{fig:SPP-S-Params}
     \end{subfigure}
     \hfill
     \caption{\ref{fig:SPP-Waveguide} Schematic drawing of the SPP waveguide test structures integrated with CPW probe pad transitions on InP. \ref{fig:SPP-S-Params} Numerically simulated and measured S-parameters of the SPP waveguide in the W band. There is good agreement between the measured S-Parameters. The Measured transmission coefficient at 100 GHz is around -6 db.}
\end{figure}

The waveguide transition tapers from input CPW 150 $\mu$m pitch probe pads to a small CPW section equivalent in dimensions to the CPW contacts integrated with the UTC-PD. This then includes a flared ground mode converter to convert from the CPW contacts of the UTC-PD to the SPP waveguide. Due to the mismatch in phase velocity between the CPW and SPP waveguide it is necessary to design appropriate mode converters to transfer the propagating CPW mode to the confined surface wave in the SPP waveguide\cite{Ma2014-zc}. Similar transitions have been previously developed for planar Goubau lines following a similar flared ground structure. This has also been achieved in transitioning from microstrip waveguides to SPP waveguides\cite{Liao2014-bi}. 
A test structure of the SPP waveguide that will be integrated with the UTC-PD was designed to operate in the W band. A schematic diagram of the waveguide is shown in Figure \ref{fig:SPP-Waveguide}. The waveguide tapers from input CPW 150 $\mu m$ pitch probe pads to a small CPW section equivalent in dimensions to the CPW contacts integrated with the UTC-PD. This then includes a flared ground mode converter to convert from the CPW contacts of the UTC-PD to the SPP waveguide. The numerically simulated S-Parameters are shown in Figure \ref{fig:SPP-S-Params}. and compared with measured S-Parameters with the upper band limited by the frequency extenders of the VNA. There is good agreement with the measured and modelled data. showing around 5-6 dB of propagation losses between 75-110 GHz.

\subsection*{UTC-PD integrated with metamaterial waveguide}

\begin{figure}
     \centering
     \begin{subfigure}[b]{0.45\textwidth}
         \centering
         \includegraphics[width=\textwidth]{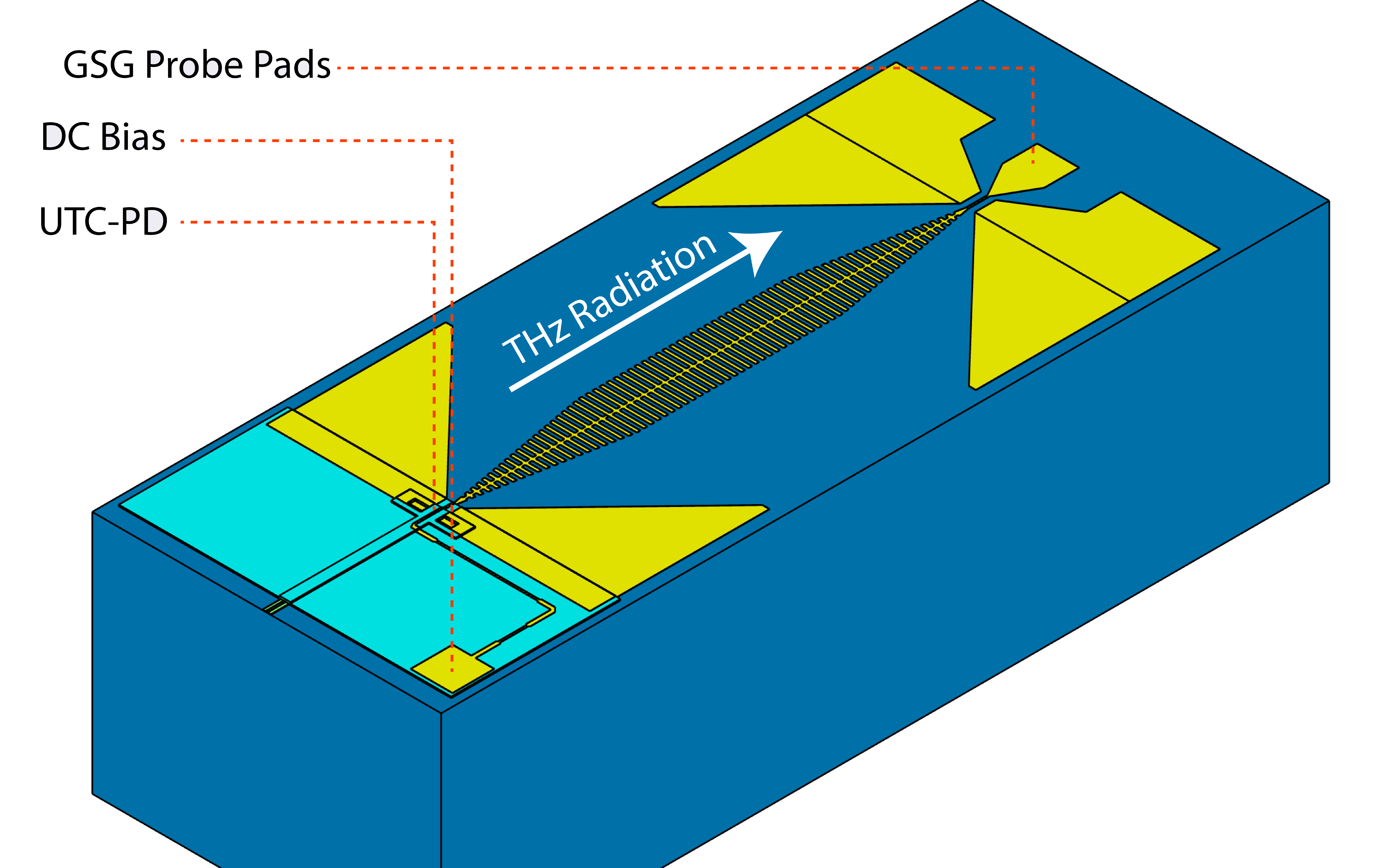}
         \caption{}
         \label{fig:SPP-UTC-PD}
     \end{subfigure}
     \hfill
     \begin{subfigure}[b]{0.5\textwidth}
         \centering
         \includegraphics[width=\textwidth]{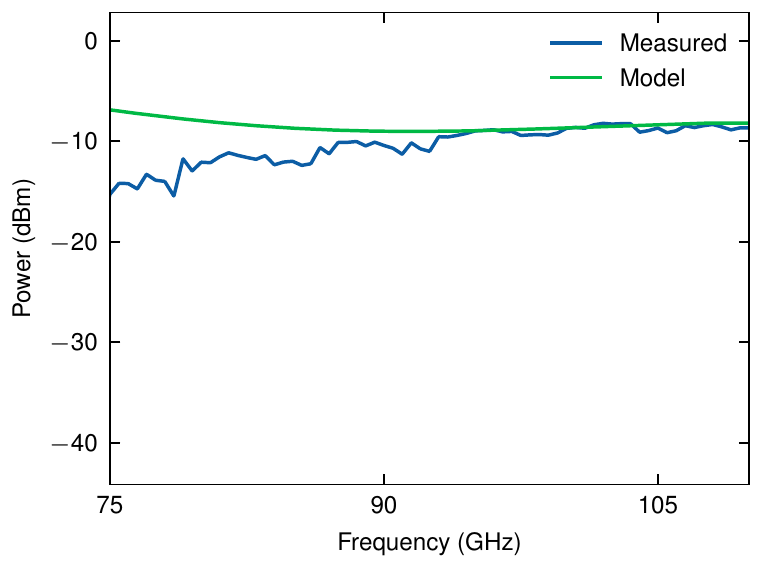}
         \caption{}
         \label{fig:SPP-UTC-Power}
     \end{subfigure}
     \hfill
     \caption{\ref{fig:SPP-UTC-PD} Schematic drawing of a UTC-PD integrated with a SPP waveguide, terminated with CPW probe pads. A bias line is added to the opposite side of the SPP waveguide with a planar stepped impedance filter to block the generated THz power from passing through this line. \ref{fig:SPP-UTC-Power} Output RF power measured using an RF power meter connected to the device via an ACP probe. A maximum output power of -9 dBm is recorded at 110 GHz, the losses of the probe setup have been measured and de-embedded from the power measurements. }
\end{figure}

A section of SPP waveguide was integrated with a UTC-PD and terminated with a transition back to CPW contact pads for output coupling of the THz power to an air co-planar (ACP) probe. To isolate the IF signal as part of the receiver and provide independent bias to both the Rx and Tx UTC-PD a low pass filter was integrated through a stepped impedance microstrip inductor to the bias line of the UTC-PD as the flared grounds of the SPP waveguide transition form an effective DC block at the RF input. A schematic diagram of the UTC-PD integrated with SPP waveguide is shown in Figure \ref{fig:SPP-UTC-PD}. The length of the taper and the flared grounds were optimised to maximise power transmitted in the 75-110 GHz band. The taper serves two functions, the first is to compensate for the mismatch in momentum between the guided CPW mode at the UTC-PD electrodes and the SPP waveguide. In addition to this the taper provides a method of impedance tuning to help match the complex impedance of the UTC-PD to the transmission line characteristic impedance and 50 Ohm termination of the waveguide.

\begin{figure}[hb]
\centering
\includegraphics[width=\linewidth]{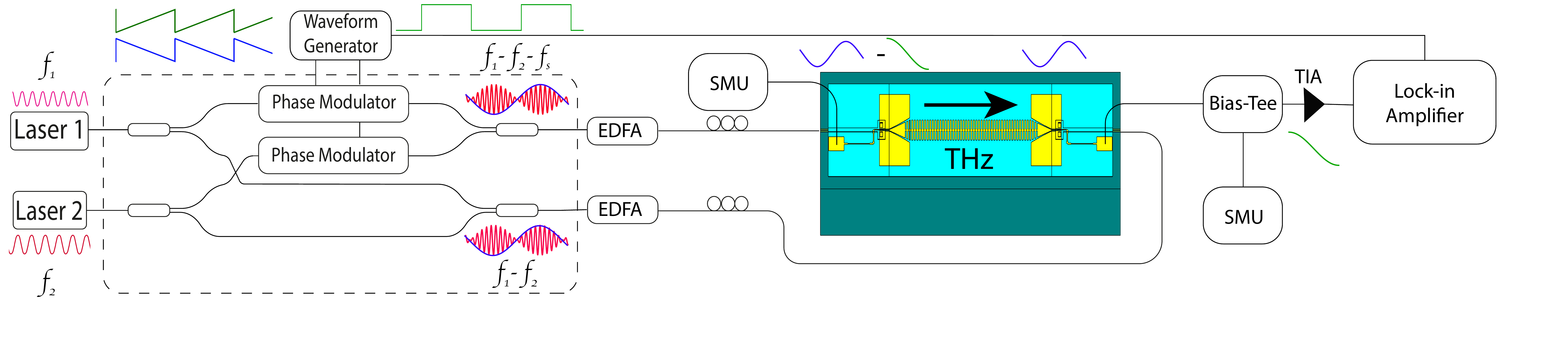}
\caption{Schematic drawing of the experimental arrangement for evaluating the dynamic range of the on-chip spectrometer system. The optical heterodyne generation system is the same as is used for the free space experiments. The optical drive signals are delivered to the photodiodes via tapered lensed fibres. The DC bias for the Tx and Rx photodiodes are applied using single pin probes. The down converted THz signal is then recorded using a lock in amplifier with a 100 ms time constant}
\label{fig:LOC_Schematic}
\end{figure}
The output power of the SPP waveguide UTC-PD was measured using an air co-planar (ACP) probe and a W band power meter. The UTC-PD was driven with a heterodyne optical signal from two free running lasers. One of the laser tones was tuned across the waveguide probe frequency bands and the output power and photocurrent were recorded. The power meter was connected to the probe by a W band semi-flexible waveguide. The S-parameters of the semi-flexible waveguide were measured and de-embedded from the power measurements along with the ACP probe losses. Figure \ref{fig:SPP-UTC-Power} shows the measured output power on-chip in the W band frequency range for a photocurrent of 10 mA.
There is good agreement between the measured output power and the numerical model showing improving output as the frequency increases. To demonstrate a combined transmitter and receiver integrated on-chip two UTC-PDs were integrated with a section of SPP waveguide and integrated mode converters shown in the schematic diagram in Figure \ref{fig:LOC_Schematic}.
The same bias line filters were used from the single UTC-PD integrated with metamaterial waveguide to provide DC bias to the photodiodes and extract the down converted IF signal from the receiver PD. 

 \begin{figure}[ht]
    \centering
    \includegraphics[width=0.6\linewidth]{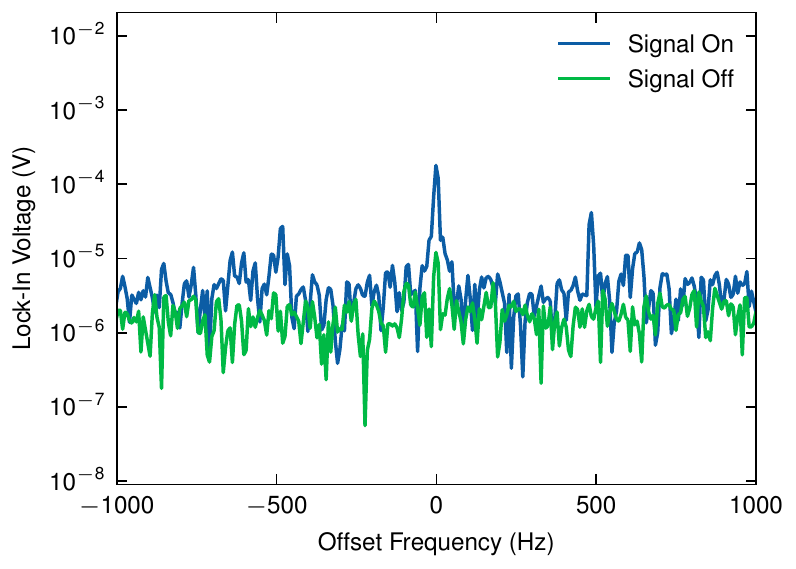}
    \caption{Fourier transform spectrum of the IF signal with the transmitter on and off. A residual peak is observed for the off case which is attributed to stray light from the transmitter photodiode crossing the chip and beating with the unshifted signals in the receiver photodiode.}
    \label{fig:IF_SPectrum}
\end{figure}

\subsection*{Optimum mixing parameters on chip }

\begin{figure}[ht]
    \centering
    \includegraphics[width=0.6\linewidth]{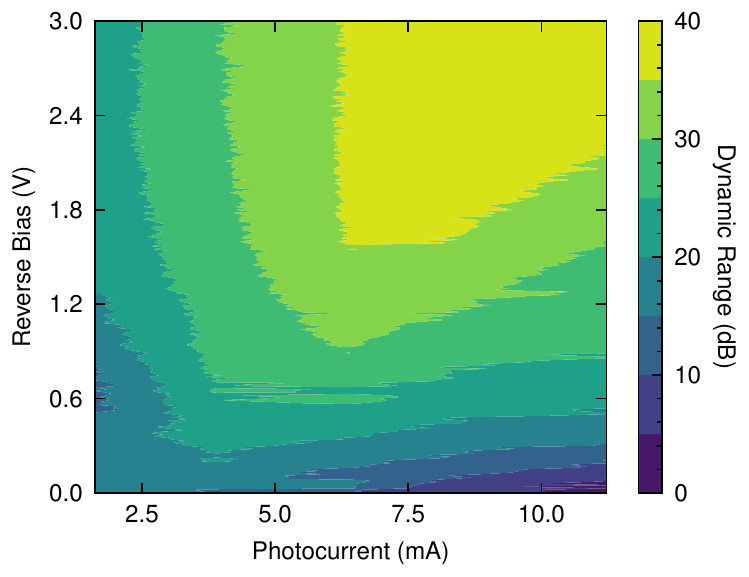}
    \caption{Measurement of the optimum mixing conditions for the on-chip spectrometer. A similar optimum is observed when compared to the equivalent free space system where the peak dynamic range is observed at high optical drive powers and high reverse biases. }
    \label{fig:IF_Response_LOC}
\end{figure}

As with the free space spectrometer based on UTC-PDs the bias and optical power were swept on the receiver UTC-PD to determine the parameters that yielded the maximum dynamic range.
The RMS noise floor was measured in two conditions, the first conditions was with the Rx optical power. A reduced signal at 20 kHz was still observed when the transmitter PD was on, indicating that either light from the Tx UTC-PD was able to leak into the receiver UTC-PD generating a beat note at the 20 kHz IF frequency or a beat note generated in the TX PD from parasitic amplitude modulation was coupled to the receiver circuit. This had the effect of limiting the overall dynamic range of the system shown in the spectrum in Figure \ref{fig:IF_SPectrum}. The second case was with the transmitter optical power off, it was found that the RMS noise is dominated by the coupling of the stray light from the transmitter photodiode across the chip to the receiver photodiode, so this level was measured and used as the RMS noise floor for dynamic range calculations. To mitigate this issue in the future the next iterations of the chip could address this by laterally offsetting the two PDs using an S bend SPP waveguide. Future iterations of the chip could address this by offsetting the two PDs using an S bend SPP waveguide. Future iterations of the chip could address this by offsetting the two PDs using an S bend SPP waveguide. A colourmap of the optimum point of operation for the on-chip spectrometer is shown in Figure \ref{fig:IF_Response_LOC}. As with the free space equivalent system the optimum point of operation for the receiver is at high reverse bias and high optical LO power. 

\subsection*{On chip dynamic range}
\begin{figure}[ht]
    \centering
    \includegraphics[width=0.6\linewidth]{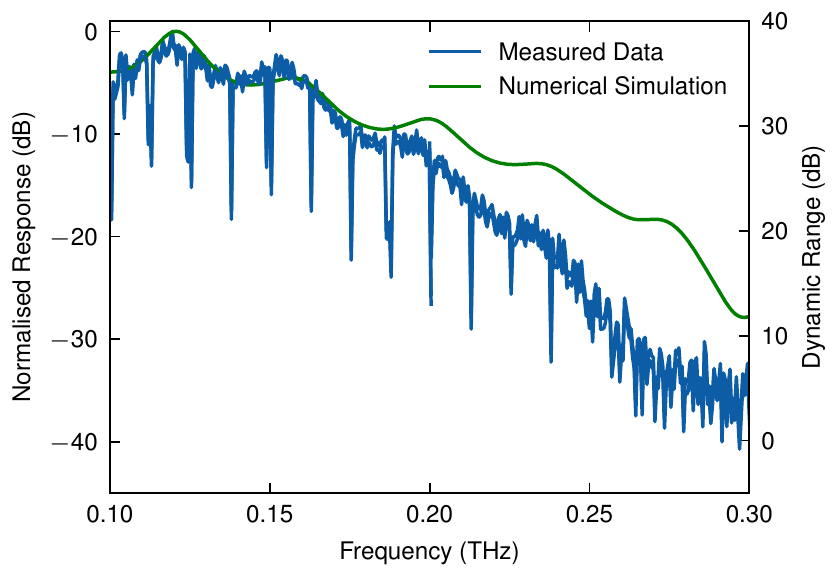}
    \caption{Measurement of dynamic range spectrum of the on-chip spectrometer. The bias and optical power conditions are set to the measured optimum values. A 20 kHz IF is used with a lock in amplifier time constant of 100 ms.}
    \label{fig:Dynamic_Range_LOC}
\end{figure}

The heterodyne frequency was swept from 0.1-0.3 THz to record the dynamic range spectrum of the on-chip spectrometer. The RMS noise was measured across the same range using the technique described in the optimum on-chip mixing experiment.  Figure \ref{fig:Dynamic_Range_LOC} shows the dynamic range of the spectrometer across the measured bandwidth. This is compared to the full wave model response of the system based on simulated power generated in a transmitter UTC-PD with a photocurrent of 12 mA. The sharp regularly spaced dips in the spectrum as with the free space spectrometer measurements are due to laser mode hops as the tuneable laser wavelength is swept. There is also a periodic ripple present in both the measured and numerically simulated data indicating the presence of a standing wave on the transmission line caused by mismatch between source and load impedances and the characteristic impedance of the transmission line. A peak dynamic range of 40 dB is observed around 120 GHz, this corresponds with the numerical simulations showing that the most power is delivered to the Rx UTC-PD at this frequency. The peak dynamic range is 20 dB down from the maximum observed in the free space system. This reduction is likely due to several factors, the principle being impedance mismatches between the source and load on the transmission line. In addition to this the transmitted power is also attenuated by the waveguide through ohmic and radiative losses. Tuning of the spoof plasma frequency of the waveguide to a higher cut off frequency should help to alleviate ohmic losses, potentially extending the bandwidth of the system beyond 300 GHz. 

\section*{Conclusions}
We have demonstrated the first proof of concept example of a CW THz spectrometer using UTC-PDs. Monolithic integration of the THz source and receiver allows for a compact sensing platform to be developed and can be further reduced in size through monolithic integration of the lasers and optical frequency shifting network. We also demonstrate for the first time, integration of a UTC-PD with a metamaterial waveguide which confines the THz radiation to a subwavelength grooved interface. The on chip spectrometer has a bandwidth of 0.3 THz and peak dynamic range of 40 dB. Coupling of single metamaterial elements such as inductive capacitive (LC) high Q resonators which could act as either functionalised bio sensors or regions or a near field probe for compact near field imaging and strong interactions with an absorbing sample could extend the functionality of the platform further.

\section*{Methods}

\subsection*{Numerical Methods}
Full wave numerical simulations of the waveguides and devices were carried out using CST microwave studio frequency domain solver. A surface impedance model for gold was used to account for the metal losses in the waveguide\cite{Lucyszyn2007-cq}. The substrate was simulated using a lossy dielectric model. For the photodiode integrated devices an equivalent circuit model44 for the UTC-PD was used. The parameters of the equivalent circuit were optimised to fit the device impedance of the receiver UTC-PD when it is driven at optimum conditions for maximum dynamic range which were obtained from the free space spectrometer measurements. A numerical optimisation function was used within CST studio to vary the CPW-SPP waveguide transition length and ground flaring to maximise simulated power delivered to the receiver photodiode. 
\subsection*{Device Fabrication}
The spoof plasmon waveguides were fabricated on semi insulating indium phosphide via sputtering of titanium gold onto the substrate. 
The UTC-PDS were grown using solid source MBE and then fabricated using photolithography to define the diode ridge structure and optical waveguides followed by passivation using silicon oxynitride finally the waveguides were deposited using sputtering of titanium gold. The samples were then cleaved to reveal the optical facets of the transmitter and receiver photodiodes. 
\subsection*{On Wafer Network Analysis}
A Keysight PNAX N277A network analyser with VDI W band frequency extenders were used to measure network parameters. An on chip 2 port SOLT calibration was carried to de-embed the effects of the RF probes. The impedance of the UTC-PD was measured using the same network analyser from 100 MHz to 67 GHz and using the W band extenders, a 1 port SOL calibration was used to de-embed the probes.
\subsection*{Optical Heterodyne Generation}
The input optical heterodyne signal for the free space measurements and the on-chip measurements was generated using two lasers, a fixed wavelength laser (Rio Orion series) and a tuneable external cavity laser (Santec TSL 710) a polarization controller was placed after each laser to set both laser polarizations parallel to each other. The two laser tones were combined using a 50:50 fibre coupler. The optical frequency shifting was carried out using a pair of electro-optic phase modulators and a differential sawtooth drive waveform to generate a Serrodyne frequency shift40. Two phase modulators were used to translate the optical frequency to the second harmonic of the sawtooth drive frequency and a 4V$\pi$ peak to peak drive voltage was used to shift the frequency further to the 4th harmonic\cite{Kim2013-by}. An EDFA was used after the frequency translation for both the Tx and Rx optical paths to increase the optical power. For the free space spectrometer, packaged antenna integrated UTC-PD modules were used as the THz source and receiver and the beam path was formed by 4 50.8 mm diameter off axis parabolic mirrors with a total path length of 90 cm. For the on-chip spectrometer, a lensed fibre with a 2.5 $\mu$m spot size was used to deliver the optical drive signals to each photodiode. DC bias was supplied via single pin probes placed on the contact pads of each photodiode. The IF signal was extracted via the single pin probes on the receiver photodiode, the DC bias and RF were separated using a bias-tee, the IF signal was then amplified using a DPLCA Femto variable gain preamplifier with gain set to 104 V/A. Finally, the IF signal was recorded on a Zurich instruments MFLI with a 100 ms time constant. 
\bibliography{paperpile}

\begin{thebibliography}{10}
\urlstyle{rm}
\expandafter\ifx\csname url\endcsname\relax
  \def\url#1{\texttt{#1}}\fi
\expandafter\ifx\csname urlprefix\endcsname\relax\def\urlprefix{URL }\fi
\expandafter\ifx\csname doiprefix\endcsname\relax\def\doiprefix{DOI: }\fi
\providecommand{\bibinfo}[2]{#2}
\providecommand{\eprint}[2][]{\url{#2}}

\bibitem{Dhillon2017-jk}
\bibinfo{author}{Dhillon, S.~S.} \emph{et~al.}
\newblock \bibinfo{title}{The 2017 terahertz science and technology roadmap}
  (\bibinfo{year}{2017}).

\bibitem{Globus2013-vw}
\bibinfo{author}{Globus, T.} \emph{et~al.}
\newblock \bibinfo{journal}{\bibinfo{title}{Highly resolved {Sub-Terahertz}
  vibrational spectroscopy of biological macromolecules and cells}}.
\newblock {\emph{\JournalTitle{IEEE Sens. J.}}} \textbf{\bibinfo{volume}{13}},
  \bibinfo{pages}{72--79} (\bibinfo{year}{2013}).

\bibitem{Motiyenko2014-qw}
\bibinfo{author}{Motiyenko, R.~A.}, \bibinfo{author}{Ilyushin, V.~V.},
  \bibinfo{author}{Drouin, B.~J.}, \bibinfo{author}{Yu, S.} \&
  \bibinfo{author}{Margul{\`e}s, L.}
\newblock \bibinfo{journal}{\bibinfo{title}{Rotational spectroscopy of
  methylamine up to 2.6 {THz}}}.
\newblock {\emph{\JournalTitle{Astron. Astrophys. Suppl. Ser.}}}
  \textbf{\bibinfo{volume}{563}}, \bibinfo{pages}{A137} (\bibinfo{year}{2014}).

\bibitem{Wirtz2018-an}
\bibinfo{author}{Wirtz, H.}, \bibinfo{author}{Sch{\"a}fer, S.},
  \bibinfo{author}{Hoberg, C.} \& \bibinfo{author}{Havenith, M.}
\newblock \bibinfo{journal}{\bibinfo{title}{Differences in hydration structure
  around hydrophobic and hydrophilic model peptides probed by {THz}
  spectroscopy}}.
\newblock {\emph{\JournalTitle{J. Infrared Millim. Terahertz Waves}}}
  \textbf{\bibinfo{volume}{39}}, \bibinfo{pages}{816--827}
  (\bibinfo{year}{2018}).

\bibitem{Esser2018-jt}
\bibinfo{author}{Esser, A.} \emph{et~al.}
\newblock \bibinfo{journal}{\bibinfo{title}{Hydrophilic solvation dominates the
  terahertz fingerprint of amino acids in water}}.
\newblock {\emph{\JournalTitle{J. Phys. Chem. B}}}
  \textbf{\bibinfo{volume}{122}}, \bibinfo{pages}{1453--1459}
  (\bibinfo{year}{2018}).

\bibitem{Swearer2018-bz}
\bibinfo{author}{Swearer, D.~F.} \emph{et~al.}
\newblock \bibinfo{journal}{\bibinfo{title}{Monitoring chemical reactions with
  terahertz rotational spectroscopy}}.
\newblock {\emph{\JournalTitle{ACS Photonics}}} \textbf{\bibinfo{volume}{5}},
  \bibinfo{pages}{3097--3106} (\bibinfo{year}{2018}).

\bibitem{Matmon2016-xb}
\bibinfo{author}{Matmon, G.}, \bibinfo{author}{Lynch, S.~A.},
  \bibinfo{author}{Rosenbaum, T.~F.}, \bibinfo{author}{Fisher, A.~J.} \&
  \bibinfo{author}{Aeppli, G.}
\newblock \bibinfo{journal}{\bibinfo{title}{Optical response from terahertz to
  visible light of electronuclear transitions in {LiYF4:Ho3+}}}.
\newblock {\emph{\JournalTitle{Phys. Rev. B: Condens. Matter Mater. Phys.}}}
  \textbf{\bibinfo{volume}{94}}, \bibinfo{pages}{17--19}
  (\bibinfo{year}{2016}).

\bibitem{Heyman1998-rf}
\bibinfo{author}{Heyman, J.~N.}, \bibinfo{author}{Kersting, R.} \&
  \bibinfo{author}{Unterrainer, K.}
\newblock \bibinfo{journal}{\bibinfo{title}{Time-domain measurement of
  intersubband oscillations in a quantum well}}.
\newblock {\emph{\JournalTitle{Appl. Phys. Lett.}}}
  \textbf{\bibinfo{volume}{72}}, \bibinfo{pages}{644--646}
  (\bibinfo{year}{1998}).

\bibitem{Stanze2011-ay}
\bibinfo{author}{Stanze, D.} \emph{et~al.}
\newblock \bibinfo{journal}{\bibinfo{title}{Compact cw terahertz spectrometer
  pumped at 1.5 $\mu$m wavelength}}.
\newblock {\emph{\JournalTitle{J. Infrared Millim. Terahertz Waves}}}
  \textbf{\bibinfo{volume}{32}}, \bibinfo{pages}{225--232}
  (\bibinfo{year}{2011}).

\bibitem{Ishibashi1997-gn}
\bibinfo{author}{Ishibashi, T.} \emph{et~al.}
\newblock \bibinfo{title}{{Uni-Traveling-Carrier} photodiodes}.
\newblock In \emph{\bibinfo{booktitle}{Ultrafast Electronics and
  Optoelectronics}}, vol.~\bibinfo{volume}{2}, \bibinfo{pages}{808--809}
  (\bibinfo{publisher}{OSA}, \bibinfo{address}{Washington, D.C.},
  \bibinfo{year}{1997}).

\bibitem{Nagatsuma2010-ez}
\bibinfo{author}{Nagatsuma, T.} \emph{et~al.}
\newblock \bibinfo{journal}{\bibinfo{title}{Continuous-wave terahertz
  spectroscopy system based on photodiodes}}.
\newblock {\emph{\JournalTitle{PIERS Online}}} \textbf{\bibinfo{volume}{6}},
  \bibinfo{pages}{390--394} (\bibinfo{year}{2010}).

\bibitem{Rouvalis2011-fk}
\bibinfo{author}{Rouvalis, E.}, \bibinfo{author}{Fice, M.~J.},
  \bibinfo{author}{Renaud, C.~C.} \& \bibinfo{author}{Seeds, A.~J.}
\newblock \bibinfo{journal}{\bibinfo{title}{Optoelectronic detection of
  millimetre-wave signals with travelling-wave uni-travelling carrier
  photodiodes}}.
\newblock {\emph{\JournalTitle{Opt. Express}}} \textbf{\bibinfo{volume}{19}},
  \bibinfo{pages}{2079} (\bibinfo{year}{2011}).

\bibitem{Rouvalis2012-zw}
\bibinfo{author}{Rouvalis, E.}, \bibinfo{author}{Fice, M.~J.},
  \bibinfo{author}{Renaud, C.~C.} \& \bibinfo{author}{Seeds, A.~J.}
\newblock \bibinfo{journal}{\bibinfo{title}{{Millimeter-Wave} optoelectronic
  mixers based on {Uni-Traveling} carrier photodiodes}}.
\newblock {\emph{\JournalTitle{IEEE Trans. Microw. Theory Tech.}}}
  \textbf{\bibinfo{volume}{60}}, \bibinfo{pages}{686--691}
  (\bibinfo{year}{2012}).

\bibitem{TOPTICA_Photonics2021-ep}
\bibinfo{author}{{TOPTICA Photonics}}.
\newblock \bibinfo{title}{{TeraScan} {Frequency-Domain} terahertz platform}.
\newblock
  \bibinfo{howpublished}{\url{https://www.toptica.com/products/terahertz-systems/frequency-domain/terascan/}}
  (\bibinfo{year}{2021}).

\bibitem{Naftaly2013-is}
\bibinfo{author}{Naftaly, M.}
\newblock \bibinfo{journal}{\bibinfo{title}{Metrology issues and solutions in
  {THz} {Time-Domain} spectroscopy: Noise, errors, calibration}}.
\newblock {\emph{\JournalTitle{IEEE Sens. J.}}} \textbf{\bibinfo{volume}{13}},
  \bibinfo{pages}{8--17} (\bibinfo{year}{2013}).

\bibitem{Byrne2008-zo}
\bibinfo{author}{Byrne, M.~B.} \emph{et~al.}
\newblock \bibinfo{journal}{\bibinfo{title}{Terahertz vibrational absorption
  spectroscopy using microstrip-line waveguides}}.
\newblock {\emph{\JournalTitle{Appl. Phys. Lett.}}}
  \textbf{\bibinfo{volume}{93}}, \bibinfo{pages}{182904}
  (\bibinfo{year}{2008}).

\bibitem{Auston1980-ao}
\bibinfo{author}{Auston, D.~H.}, \bibinfo{author}{Johnson, A.~M.},
  \bibinfo{author}{Smith, P.~R.} \& \bibinfo{author}{Bean, J.~C.}
\newblock \bibinfo{journal}{\bibinfo{title}{Picosecond optoelectronic
  detection, sampling, and correlation measurements in amorphous
  semiconductors}}.
\newblock {\emph{\JournalTitle{Appl. Phys. Lett.}}}
  \textbf{\bibinfo{volume}{37}}, \bibinfo{pages}{371--373}
  (\bibinfo{year}{1980}).

\bibitem{Grischkowsky2000-fc}
\bibinfo{author}{Grischkowsky, D.~R.}
\newblock \bibinfo{journal}{\bibinfo{title}{Optoelectronic characterization of
  transmission lines and waveguides by terahertz time-domain spectroscopy}}.
\newblock {\emph{\JournalTitle{IEEE J. Sel. Top. Quantum Electron.}}}
  \textbf{\bibinfo{volume}{6}}, \bibinfo{pages}{1122--1135}
  (\bibinfo{year}{2000}).

\bibitem{Wood2006-at}
\bibinfo{author}{Wood, C.} \emph{et~al.}
\newblock \bibinfo{journal}{\bibinfo{title}{On-chip photoconductive excitation
  and detection of pulsed terahertz radiation at cryogenic temperatures}}.
\newblock {\emph{\JournalTitle{Appl. Phys. Lett.}}}
  \textbf{\bibinfo{volume}{88}}, \bibinfo{pages}{142103}
  (\bibinfo{year}{2006}).

\bibitem{Park2020-nd}
\bibinfo{author}{Park, S.~J.} \& \bibinfo{author}{Cunningham, J.}
\newblock \bibinfo{journal}{\bibinfo{title}{Determination of permittivity of
  dielectric analytes in the terahertz frequency range using split ring
  resonator elements integrated with {On-Chip} waveguide}}.
\newblock {\emph{\JournalTitle{Sensors}}} \textbf{\bibinfo{volume}{20}}
  (\bibinfo{year}{2020}).

\bibitem{Swithenbank2017-dd}
\bibinfo{author}{Swithenbank, M.} \emph{et~al.}
\newblock \bibinfo{journal}{\bibinfo{title}{{On-Chip} {Terahertz-Frequency}
  measurements of liquids}}.
\newblock {\emph{\JournalTitle{Anal. Chem.}}} \textbf{\bibinfo{volume}{89}},
  \bibinfo{pages}{7981--7987} (\bibinfo{year}{2017}).

\bibitem{Wood2013-ip}
\bibinfo{author}{Wood, C.~D.} \emph{et~al.}
\newblock \bibinfo{journal}{\bibinfo{title}{On-chip terahertz spectroscopic
  techniques for measuring mesoscopic quantum systems}}.
\newblock {\emph{\JournalTitle{Rev. Sci. Instrum.}}}
  \textbf{\bibinfo{volume}{84}}, \bibinfo{pages}{085101}
  (\bibinfo{year}{2013}).

\bibitem{Goubau1951-nc}
\bibinfo{author}{Goubau, G.}
\newblock \bibinfo{journal}{\bibinfo{title}{{Single-Conductor} {Surface-Wave}
  transmission lines}}.
\newblock {\emph{\JournalTitle{Proceedings of the IRE}}}
  \textbf{\bibinfo{volume}{39}}, \bibinfo{pages}{619--624}
  (\bibinfo{year}{1951}).

\bibitem{Dazhang2009-lm}
\bibinfo{author}{Dazhang, L.} \emph{et~al.}
\newblock \bibinfo{journal}{\bibinfo{title}{On-chip terahertz goubau-line
  waveguides with integrated photoconductive emitters and mode-discriminating
  detectors}}.
\newblock {\emph{\JournalTitle{Appl. Phys. Lett.}}}
  \textbf{\bibinfo{volume}{95}}, \bibinfo{pages}{092903}
  (\bibinfo{year}{2009}).

\bibitem{Akalin2019-oz}
\bibinfo{author}{Akalin, T.}, \bibinfo{author}{Chahadih, A.},
  \bibinfo{author}{Ghaddar, A.} \& \bibinfo{author}{T{\"\i}rer, I.}
\newblock \bibinfo{title}{Combined {UTC-PD} integrated on-chip {THz} near field
  microscopy with coupled planar goubau lines}.
\newblock In \emph{\bibinfo{booktitle}{2019 44th International Conference on
  Infrared, Millimeter, and Terahertz Waves ({IRMMW-THz})}},
  \bibinfo{pages}{1--2} (\bibinfo{year}{2019}).

\bibitem{Russell2013-xu}
\bibinfo{author}{Russell, C.} \emph{et~al.}
\newblock \bibinfo{title}{Optimization and application of on-chip terahertz
  goubau lines}.
\newblock In \emph{\bibinfo{booktitle}{2013 38th International Conference on
  Infrared, Millimeter, and Terahertz Waves ({IRMMW-THz})}},
  \bibinfo{pages}{1--2} (\bibinfo{year}{2013}).

\bibitem{Goubau1950-zq}
\bibinfo{author}{Goubau, G.}
\newblock \bibinfo{journal}{\bibinfo{title}{Surface waves and their application
  to transmission lines}}.
\newblock {\emph{\JournalTitle{J. Appl. Phys.}}} \textbf{\bibinfo{volume}{21}},
  \bibinfo{pages}{1119--1128} (\bibinfo{year}{1950}).

\bibitem{Barlow1953-xx}
\bibinfo{author}{Barlow, H.~M.} \& \bibinfo{author}{Cullen, A.~L.}
\newblock \bibinfo{journal}{\bibinfo{title}{Surface waves}}.
\newblock {\emph{\JournalTitle{Proceedings of the IEE - Part III: Radio and
  Communication Engineering}}} \textbf{\bibinfo{volume}{100}},
  \bibinfo{pages}{329--341} (\bibinfo{year}{1953}).

\bibitem{Pendry2004-zx}
\bibinfo{author}{Pendry, J.~B.}, \bibinfo{author}{Mart{\'\i}n-Moreno, L.} \&
  \bibinfo{author}{Garcia-Vidal, F.~J.}
\newblock \bibinfo{journal}{\bibinfo{title}{Mimicking surface plasmons with
  structured surfaces}}.
\newblock {\emph{\JournalTitle{Science}}} \textbf{\bibinfo{volume}{305}},
  \bibinfo{pages}{847--848} (\bibinfo{year}{2004}).

\bibitem{Shen2013-is}
\bibinfo{author}{Shen, X.}, \bibinfo{author}{Cui, T.~J.},
  \bibinfo{author}{Martin-Cano, D.} \& \bibinfo{author}{Garcia-Vidal, F.~J.}
\newblock \bibinfo{journal}{\bibinfo{title}{Conformal surface plasmons
  propagating on ultrathin and flexible films}}.
\newblock {\emph{\JournalTitle{Proc. Natl. Acad. Sci. U. S. A.}}}
  \textbf{\bibinfo{volume}{110}}, \bibinfo{pages}{40--45}
  (\bibinfo{year}{2013}).

\bibitem{Williams2008-pr}
\bibinfo{author}{Williams, C.~R.} \emph{et~al.}
\newblock \bibinfo{journal}{\bibinfo{title}{Highly confined guiding of
  terahertz surface plasmon polaritons on structured metal surfaces}}.
\newblock {\emph{\JournalTitle{Nat. Photonics}}} \textbf{\bibinfo{volume}{2}},
  \bibinfo{pages}{175--179} (\bibinfo{year}{2008}).

\bibitem{Nagel2002-tf}
\bibinfo{author}{Nagel, M.} \emph{et~al.}
\newblock \bibinfo{journal}{\bibinfo{title}{Integrated {THz} technology for
  label-free genetic diagnostics}}.
\newblock {\emph{\JournalTitle{Appl. Phys. Lett.}}}
  \textbf{\bibinfo{volume}{80}}, \bibinfo{pages}{154--156}
  (\bibinfo{year}{2002}).

\bibitem{Lamponi2013-sc}
\bibinfo{author}{Lamponi, M.} \emph{et~al.}
\newblock \bibinfo{title}{Tunable {InP} photonic integrated circuit for
  millimeter wave generation}.
\newblock In \emph{\bibinfo{booktitle}{2013 International Conference on Indium
  Phosphide and Related Materials ({IPRM})}}, \bibinfo{pages}{1--2}
  (\bibinfo{publisher}{IEEE}, \bibinfo{year}{2013}).

\bibitem{Lo2017-wb}
\bibinfo{author}{Lo, M.-C.} \emph{et~al.}
\newblock \bibinfo{title}{Foundry-fabricated {DFB} laser with waveguide
  crossing for self-heterodyne terahertz spectrometer}.
\newblock In \emph{\bibinfo{booktitle}{2017 Progress in Electromagnetics
  Research Symposium - Fall ({PIERS} - {FALL})}}, \bibinfo{pages}{1520--1523}
  (\bibinfo{year}{2017}).

\bibitem{Nishi2014-un}
\bibinfo{author}{Nishi, H.} \emph{et~al.}
\newblock \bibinfo{title}{Si photonic integrated circuit for a compact and
  stable {CW-THz} spectrometer}.
\newblock In \emph{\bibinfo{booktitle}{11th International Conference on Group
  {IV} Photonics ({GFP})}}, \bibinfo{pages}{223--224} (\bibinfo{year}{2014}).

\bibitem{Hisatake2013-tp}
\bibinfo{author}{Hisatake, S.} \emph{et~al.}
\newblock \bibinfo{journal}{\bibinfo{title}{Phase-sensitive terahertz
  self-heterodyne system based on photodiode and low-temperature-grown {GaAs}
  photoconductor at 1.55 $\mu{\rm m}$}}.
\newblock {\emph{\JournalTitle{IEEE Sens. J.}}} \textbf{\bibinfo{volume}{13}},
  \bibinfo{pages}{31--36} (\bibinfo{year}{2013}).

\bibitem{Kim2013-by}
\bibinfo{author}{Kim, J.~Y.}, \bibinfo{author}{Song, H.~J.},
  \bibinfo{author}{Ajito, K.}, \bibinfo{author}{Yaita, M.} \&
  \bibinfo{author}{Kukutsu, N.}
\newblock \bibinfo{journal}{\bibinfo{title}{Continuous-wave {THz} homodyne
  spectroscopy and imaging system with electro-optical phase modulation for
  high dynamic range}}.
\newblock {\emph{\JournalTitle{IEEE Transactions on Terahertz Science and
  Technology}}} \textbf{\bibinfo{volume}{3}}, \bibinfo{pages}{158--164}
  (\bibinfo{year}{2013}).

\bibitem{Fushimi2004-ah}
\bibinfo{author}{Fushimi, H.}, \bibinfo{author}{Furuta, T.},
  \bibinfo{author}{Ishibashi, T.} \& \bibinfo{author}{Ito, H.}
\newblock \bibinfo{journal}{\bibinfo{title}{Photoresponse nonlinearity of a
  uni-traveling-carrier photodiode and its application to optoelectronic
  millimeter-wave mixing in 60 {GHz} band}}.
\newblock {\emph{\JournalTitle{Japanese Journal of Applied Physics, Part 2:
  Letters}}} \textbf{\bibinfo{volume}{43}}, \bibinfo{pages}{L966--L968}
  (\bibinfo{year}{2004}).

\bibitem{Hisatake2014-np}
\bibinfo{author}{Hisatake, S.}, \bibinfo{author}{Kim, J.-Y.},
  \bibinfo{author}{Ajito, K.} \& \bibinfo{author}{Nagatsuma, T.}
\newblock \bibinfo{journal}{\bibinfo{title}{{Self-Heterodyne} spectrometer
  using {Uni-Traveling-Carrier} photodiodes for {Terahertz-Wave} generators and
  optoelectronic mixers}}.
\newblock {\emph{\JournalTitle{J. Lightwave Technol.}}}
  \textbf{\bibinfo{volume}{32}}, \bibinfo{pages}{3683--3689}
  (\bibinfo{year}{2014}).

\bibitem{Johnson1988-oe}
\bibinfo{author}{Johnson, L.~M.} \& \bibinfo{author}{Cox, C.~H.}
\newblock \bibinfo{journal}{\bibinfo{title}{Serrodyne optical frequency
  translation with high sideband suppression}}.
\newblock {\emph{\JournalTitle{J. Lightwave Technol.}}}
  \textbf{\bibinfo{volume}{6}}, \bibinfo{pages}{109--112}
  (\bibinfo{year}{1988}).

\bibitem{Poberezhskiy2005-js}
\bibinfo{author}{Poberezhskiy, I.~Y.}, \bibinfo{author}{Bortnik, B.},
  \bibinfo{author}{Chou, J.}, \bibinfo{author}{Jalali, B.} \&
  \bibinfo{author}{Fetterman, H.~R.}
\newblock \bibinfo{journal}{\bibinfo{title}{Serrodyne frequency translation of
  continuous optical signals using ultrawide-band electrical sawtooth
  waveforms}}.
\newblock {\emph{\JournalTitle{IEEE J. Quantum Electron.}}}
  \textbf{\bibinfo{volume}{41}}, \bibinfo{pages}{1533--1539}
  (\bibinfo{year}{2005}).

\bibitem{Naftaly2009-ef}
\bibinfo{author}{Naftaly, M.} \& \bibinfo{author}{Dudley, R.}
\newblock \bibinfo{journal}{\bibinfo{title}{Methodologies for determining the
  dynamic ranges and signal-to-noise ratios of terahertz time-domain
  spectrometers}}.
\newblock {\emph{\JournalTitle{Opt. Lett.}}} \textbf{\bibinfo{volume}{34}},
  \bibinfo{pages}{1213--1215} (\bibinfo{year}{2009}).

\bibitem{Seddon2022-mz}
\bibinfo{author}{Seddon, J.~P.} \emph{et~al.}
\newblock \bibinfo{journal}{\bibinfo{title}{Photodiodes for terahertz
  applications}}.
\newblock {\emph{\JournalTitle{IEEE J. Sel. Top. Quantum Electron.}}}
  \textbf{\bibinfo{volume}{28}}, \bibinfo{pages}{1--12} (\bibinfo{year}{2022}).

\bibitem{Natrella2016-eu}
\bibinfo{author}{Natrella, M.} \emph{et~al.}
\newblock \bibinfo{journal}{\bibinfo{title}{Accurate equivalent circuit model
  for millimetre-wave {UTC} photodiodes}}.
\newblock {\emph{\JournalTitle{Opt. Express}}} \textbf{\bibinfo{volume}{24}},
  \bibinfo{pages}{4698} (\bibinfo{year}{2016}).

\bibitem{Rutledge1983-du}
\bibinfo{author}{Rutledge, D.~B.}, \bibinfo{author}{Neikirk, D.~P.} \&
  \bibinfo{author}{{others}}.
\newblock \bibinfo{journal}{\bibinfo{title}{Integrated circuit antennas}}.
\newblock {\emph{\JournalTitle{Infrared and millimeter}}}
  \textbf{\bibinfo{volume}{10}} (\bibinfo{year}{1983}).

\bibitem{Cao2014-dh}
\bibinfo{author}{Cao, L.}, \bibinfo{author}{Grimault-Jacquin, A.-S.},
  \bibinfo{author}{Zerounian, N.} \& \bibinfo{author}{Aniel, F.}
\newblock \bibinfo{journal}{\bibinfo{title}{Design and {VNA-measurement} of
  coplanar waveguide ({CPW}) on benzocyclobutene ({BCB}) at {THz}
  frequencies}}.
\newblock {\emph{\JournalTitle{Infrared Phys. Technol.}}}
  \textbf{\bibinfo{volume}{63}}, \bibinfo{pages}{157--164}
  (\bibinfo{year}{2014}).

\bibitem{Huidobro2018-kp}
\bibinfo{author}{Huidobro, P.~A.},
  \bibinfo{author}{Fern{\'a}ndez-Dom{\'\i}nguez, A.~I.},
  \bibinfo{author}{Pendry, J.~B.}, \bibinfo{author}{Mart{\'\i}n-Moreno, L.} \&
  \bibinfo{author}{Garcia-Vidal, F.~J.}
\newblock \bibinfo{title}{Spoof surface plasmon metamaterials}
  (\bibinfo{year}{2018}).

\bibitem{Ma2014-zc}
\bibinfo{author}{Ma, H.~F.}, \bibinfo{author}{Shen, X.},
  \bibinfo{author}{Cheng, Q.}, \bibinfo{author}{Jiang, W.~X.} \&
  \bibinfo{author}{Cui, T.~J.}
\newblock \bibinfo{journal}{\bibinfo{title}{Broadband and high-efficiency
  conversion from guided waves to spoof surface plasmon polaritons}}.
\newblock {\emph{\JournalTitle{Laser and Photonics Reviews}}}
  \textbf{\bibinfo{volume}{8}}, \bibinfo{pages}{146--151}
  (\bibinfo{year}{2014}).

\bibitem{Liao2014-bi}
\bibinfo{author}{Liao, Z.}, \bibinfo{author}{Zhao, J.}, \bibinfo{author}{Pan,
  B.~C.}, \bibinfo{author}{Shen, X.~P.} \& \bibinfo{author}{Cui, T.~J.}
\newblock \bibinfo{journal}{\bibinfo{title}{Broadband transition between
  microstrip line and conformal surface plasmon waveguide}}.
\newblock {\emph{\JournalTitle{J. Phys. D Appl. Phys.}}}
  \textbf{\bibinfo{volume}{47}}, \bibinfo{pages}{315103}
  (\bibinfo{year}{2014}).

\bibitem{Lucyszyn2007-cq}
\bibinfo{author}{Lucyszyn, S.}
\newblock \bibinfo{journal}{\bibinfo{title}{Evaluating surface impedance models
  for terahertz frequencies at room temperature}}.
\newblock {\emph{\JournalTitle{Prog. Electromagn. Res. Symp.}}}
  \textbf{\bibinfo{volume}{3}}, \bibinfo{pages}{554--559}
  (\bibinfo{year}{2007}).

\end{thebibliography}

\section*{Acknowledgements}
This work was funded by EPSRC grants EP/J017671/1 Coherent Terahertz Systems (COTS)-opening up the terahertz spectrum for widespread application, EP/P021859/1 HyperTerahertz - High precision terahertz spectroscopy and microscopy and EP/L015455/1 EPSRC Centre for Doctoral Training in Integrated Photonic and Electronic Systems.

\end{document}